\documentclass{aa}

\usepackage{epsf}
\usepackage{graphics}

\begin{document}

\def\clipfig#1{\def\lbracket{[}\def\testit{#1}%
    \ifx\testit\lbracket\let\next=\optclipfig\else\let\next=\stdclipfig\fi%
    \next{#1}}
%
\newcommand {\hclipfig} [7] {\clipfig[#7]{#1}{#2}{#3}{#4}{#5}{#6}}
%
\def\usemodepsfig {\global\def\cfmode{x}\typeout{*** set clipfig to PSFIG mode ***}}
\def\usemodeepsf  {\global\def\cfmode{}\typeout{*** set clipfig to EPSF mode ***}}
\def\useunitmm    {\global\def\cfunit{x}\typeout{*** set clipfig to use mm as unit ***}}
\def\useunitcm    {\global\def\cfunit{}\typeout{*** set clipfig to use cm as unit ***}}
\def\clipfigsettings {\ifx\cfmode\empty\def\ccfmode{EPSF }\else\def\ccfmode{PSFIG }\fi%
    \ifx\cfunit\empty\def\ccfunit{cm }\else\def\ccfunit{mm }\fi%
    \typeout{*** current clipfig settings: \ccfmode mode, using \ccfunit as unit ***}}
%
%
%
%
\def\stdclipfig#1#2#3#4#5#6{\ifx\cfmode\empty%
    \let\next=\eclipfig\else\let\next=\pclipfig\fi%
    \next{#1}{#2}{#3}{#4}{#5}{#6}}
\def\optclipfig#1#2]#3#4#5#6#7#8{\ifx\cfmode\empty%
    \let\next=\ehclipfig\else\let\next=\phclipfig\fi%
    \next{#3}{#4}{#5}{#6}{#7}{#8}{#2}}
%
%
%
\newcommand {\pclipfig}[6] {\ifx\cfunit\empty%
        \psfig{figure=#1.ps,width=#2cm,bbllx=#3cm,bblly=#4cm,bburx=#5cm,%
           bbury=#6cm,clip=}\else%
        \psfig{figure=#1.ps,width=#2mm,bbllx=#3mm,bblly=#4mm,bburx=#5mm,%
           bbury=#6mm,clip=}\fi}
\newcommand {\phclipfig}[7] {\ifx\cfunit\empty%
        \hspace{#7cm}\psfig{figure=#1.ps,width=#2cm,bbllx=#3cm,bblly=#4cm,%
           bburx=#5cm,bbury=#6cm,clip=}\else%
        \hspace{#7mm}\psfig{figure=#1.ps,width=#2mm,bbllx=#3mm,bblly=#4mm,%
           bburx=#5mm,bbury=#6mm,clip=}\fi}
%
%
%
\newcommand {\eclipfig}[6]{%
  \ifx\cfunit\empty\epsfxsize=#2cm\else\epsfxsize=#2mm\fi%
  \epsfclipon\epsfverbosetrue%
  \cfcmtopspts{#3}\cfllxi=\cftempi\cfllxf=\cftempf%
  \cfcmtopspts{#4}\cfllyi=\cftempi\cfllyf=\cftempf%
  \cfcmtopspts{#5}\cfurxi=\cftempi\cfurxf=\cftempf%
  \cfcmtopspts{#6}\cfuryi=\cftempi\cfuryf=\cftempf%
  \def\cfstra{\number\cfllxi.\number\cfllxf}%
  \def\cfstrb{\number\cfllyi.\number\cfllyf}%
  \def\cfstrc{\number\cfurxi.\number\cfurxf}%
  \def\cfstrd{\number\cfuryi.\number\cfuryf}%
  \hbox{\epsfbox[{\cfstra} {\cfstrb} {\cfstrc} {\cfstrd}]{#1.ps}}}
\newcommand {\ehclipfig}[7]{%
  \ifx\cfunit\empty\epsfxsize=#2cm\else\epsfxsize=#2mm\fi%
  \epsfclipon\epsfverbosetrue%
  \cfcmtopspts{#3}\cfllxi=\cftempi\cfllxf=\cftempf%
  \cfcmtopspts{#4}\cfllyi=\cftempi\cfllyf=\cftempf%
  \cfcmtopspts{#5}\cfurxi=\cftempi\cfurxf=\cftempf%
  \cfcmtopspts{#6}\cfuryi=\cftempi\cfuryf=\cftempf%
  \def\cfstra{\number\cfllxi.\number\cfllxf}%
  \def\cfstrb{\number\cfllyi.\number\cfllyf}%
  \def\cfstrc{\number\cfurxi.\number\cfurxf}%
  \def\cfstrd{\number\cfuryi.\number\cfuryf}%
  \ifx\cfunit\empty\hspace{#7cm}\else\hspace{#7mm}\fi%
  \hbox{\epsfbox[{\cfstra} {\cfstrb} {\cfstrc} {\cfstrd}]{#1.ps}}%
  \vspace{-1mm}}
%
%
%
\newdimen\cfllxi \newdimen\cfllyi  \newdimen\cfurxi  \newdimen\cfuryi
\newdimen\cfllxf \newdimen\cfllyf  \newdimen\cfurxf  \newdimen\cfuryf
\newdimen\cftemp \newdimen\cftempi \newdimen\cftempf
\newdimen\cfpspoint \cfpspoint=1bp
%
%
%
\newcommand{\cfcmtopspts}[1]{\ifx\cfunit\empty%
  \cftemp=#1cm\else\cftemp=#1mm\fi%
  \multiply\cftemp10\divide\cftemp\cfpspoint%
  \cftempf=\cftemp\divide\cftemp10\cftempi=\cftemp\multiply\cftemp10%
  \advance\cftempf-\cftemp}
%
%
\def\cfmode{}\def\cfunit{}\clipfigsettings
%

\useunitmm


\newcommand{\lb}{$\lambda$}
\newcommand{\sm}[1]{\footnotesize {#1}}
\newcommand{\inft}{$\infty$}
\newcommand{\vlv}{$\nu L_{\rm V}$}
\newcommand{\lv}{$L_{\rm V}$}
\newcommand{\lx}{$L_{\rm x}$}
\newcommand{\lsoft}{$L_{\rm 250eV}$}
\newcommand{\lhard}{$L_{\rm 1keV}$}
\newcommand{\vlsoft}{$\nu L_{\rm 250eV}$}
\newcommand{\vlhard}{$\nu L_{\rm 1keV}$}
\newcommand{\vlir}{$\nu L_{60\mu}$}
\newcommand{\ax}{$\alpha_{\rm x}$}
\newcommand{\aopt}{$\alpha_{\rm opt}$}
\newcommand{\aoxs}{$\alpha_{\rm oxs}$}
\newcommand{\aoxh}{$\alpha_{\rm oxh}$}
\newcommand{\airhard}{$\alpha_{\rm 60\mu-hard}$}
\newcommand{\aoxsoft}{$\alpha_{\rm ox-soft}$}
\newcommand{\aio}{$\alpha_{\rm io}$}
\newcommand{\aixs}{$\alpha_{\rm ixs}$}
\newcommand{\aixh}{$\alpha_{\rm ixh}$}
\newcommand{\hb}{H$\beta_{\rm b}$}
\newcommand{\nh}{$N_{\rm H}$}
\newcommand{\nhgal}{$N_{\rm H,gal}$}
\newcommand{\nhfit}{$N_{\rm H,fit}$}
\newcommand{\ale}{$\alpha_{\rm E}$}
\newcommand{\cts}{$\rm {cts\,s}^{-1}$}
\newcommand{\pl}{$\pm$}
\newcommand{\kev}{\rm keV}
\newcommand{\rb}[1]{\raisebox{1.5ex}[-1.5ex]{#1}}
\newcommand{\ten}[2]{#1\cdot 10^{#2}}
\newcommand{\msun}{$M_{\odot}$}
\newcommand{\dM}{\dot M}
\newcommand{\dMM}{$\dot{M}/M$}
\newcommand{\dMedd}{\dot M_{\rm Edd}}
\newcommand{\kms}{km\,$\rm s^{-1}$}

\thesaurus{03(02.02.2; 11.01.2; 11.14.1; 11.09.1)}

\title{RX J1624.9+7554: A new X-ray transient AGN
}
\author{D. Grupe\inst{1, }\thanks{
Guest Observer, McDonald Observatory, University of Texas at Austin}
\and H.-C. Thomas\inst{2}
\and K.M. Leighly\inst{3}
}
\offprints{\\ D. Grupe (dgrupe@xray.mpe.mpg.de)}
\institute{MPI f\"ur extraterrestrische Physik, Giessenbachstr., 
85748 Garching, Germany
\and MPI f\"ur Astrophysik, Karl-Schwarzschildstr. 1, 85748 Garching, Germany
\and Columbia Astrophysics Laboratory, 538 West 120th St., New York, 
NY 10027, U.S.A.
}
\date{received  04 August 1999; accepted 31 August 1999}
\maketitle

\begin{abstract}
We report the discovery of a new X-ray transient AGN,  RX J1624.9+7554. 
This object appeared to be bright in the ROSAT All-Sky Survey, but had turned 
off in two pointed observations about one and a half years later.
The optical identification spectrum shows a non-emission line spectrum of a
spiral galaxy at z=0.064. 
We will discuss several hypotheses that can explain the peculiar behaviour of
this object.
\end{abstract}
\section{Introduction}
One of the results of the X-ray satellite ROSAT (Tr\"umper 1983) has been the
identification of a large number of bright soft-X-ray selected AGN (e.g 
Thomas et al. 1998, Beuermann et al. 1999, Grupe et al. 1998a, 1999a) 
during its All-Sky Survey (RASS, Voges et al. 1996). Several of
these sources have turned out to be transient in the ROSAT Position Sensitive
Proportional Counter 0.1--2.4~keV energy window (PSPC, Pfeffermann et al. 1986).  
These AGN were bright only in their 'high' state phase during the
RASS, and they had become dramatically fainter or even disappeared by
the time of their pointed observations years later.
Some prominent examples are IC 3599 (Brandt et al. 1995, Grupe
et al. 1995a), WPVS007 (Grupe et al. 1995b), or RX J0134.2--4258
(Grupe et al. 1999b, Komossa 1999).
These three examples can be considered to be representative of three different kinds of
transience. In IC 3599, an X-ray outburst (similar to that in NGC 5905, 
Bade et al. 1996) was observed during the RASS.  An associated response was
observed in the optical emission lines, as the high ionization ``coronal'' iron
lines became much fainter when the X-ray flux decreased.
WPVS007 had practically turned off by the time of its pointed observation about three years
after the RASS,  
and RX J0134.9--4258 showed a dramatic
change in its X-ray spectral shape between the RASS and the pointed observation
two years later.  

In this paper we discuss the case of RX J1624.9+7554. While many of
the other transient soft X-ray selected galaxies show nuclear emission in
their 'high' and 'low' states, e.g. WPVS007, 
and are optically identified as AGN, 
RX J1624.9+7554 shows the spectrum of a normal non-active galaxy in its
low state.

\section{\label{observe} Observations and data reduction}
RX J1624.9+7554 was observed during the RASS 
between October 07, 1990 10:13 and October 15, 1990 23:12 for a total
of 1510 s.
Photons were extracted in a circle of $250^{''}$  radius for the source and
two circles of $400^{''}$ in scan direction for the background.  
The two pointed observations were performed on January 13, 1992 
(ROR: 141820p and 141829p) and had exposures of 2373 and 2923 s,
respectively.

Optical spectra were taken in 1998 with the 2.1m telescope
at McDonald Observatory/Texas (McD).
We took two spectra, one with a
$5.1^{''}$ slit (5 min) and one with a $2.1^{''}$ slit (45 min). The weather
conditions were good, with clear skies and $2^{''}$ seeing.
The spectral resolution was about 7 \AA ~(FWHM).
Two 20-minute spectra were taken with a 1.7$^{''}$ slit 
at the 2.4m Hiltner telescope 
at the Michigan-Dartmouth-MIT Observatory (MDM) in February 1999.
Weather conditions were good with some light clouds. The spectral resolution was
approximately 4.5 \AA.
Three 2-minute R band exposures were obtained by I. Yadigaroglu at MDM in
January 1999.
Optical polarimetry was performed in March 1998 using the broad-band polarimeter
at the 2.1m at McDonald Observatory. 
A description of these measurements is given in Grupe et
al. (1998b) and references therein.

Data reduction was performed with EXSAS (Zimmermann et al. 1998)
for the ROSAT data and MIDAS and FIGARO for the optical data.  

Infrared data from the IRAS satellite were retrieved using the interactive 
XSCANPI program at the Infrared Processing and Analysis Center (IPAC)
available on the WWW. The scans were made at the optical position
of the source. 

All luminosities are calculated for $ H_0=\rm75~km~s^{-1}Mpc^{-1}$ and 
$q_0$=0.5. 

\section{\label{results} Results}
The X-ray position is $\alpha_{2000}$=16h24m56.7s, 
$\delta_{2000}=+75^{\circ}54^{'}57.5^{''}$ with a 
2$\sigma$ error radius of 14$^{''}$.
This 
coincides with the optical position measured
by the POSS scans of the US Naval Observatory leading to the position
$\alpha_{2000}$ = 16h24m56.5s,
$\delta_{2000}~=~+75^{\circ}54^{'}55.8^{''}$.  We identified RX
J1624.9+7554 as a galaxy with a redshift of z=0.0636\pl0.0005.
No other source was found inside the 2$\sigma$ error circle, either on the POSS
or on the R-image. 

\subsection{X-rays}
During the RASS observation RX J1624.9+7554 had a count rate of 0.54$\pm$0.02
$\rm cts~s^{-1}$ 
with a hardness ratio of HR=--0.20$\pm$0.04 (Thomas et al. 1998).

We performed standard spectral fits to the RASS spectrum, with all
parameters free or with neutral absorption fixed to the Galactic value
($\rm N_H~=~0.39*10^{21}~cm^{-2}$; Dickey \& Lockmann 1990).
The results are listed in Table
\ref{rass_fits}. The best fitting models are a power law with intrinsic $\rm N_H$ and
a thermal Bremsstrahlung spectrum. We cannot distinguish between these
two models on the basis of the spectral fitting.  
However, the thermal Bremsstrahlung is ruled out because the emission
region size implied by the variability between the RASS and the
pointed observation is small enough that the assumption of optically
thin gas is violated (see e.g Elvis et al. 1991).
Furthermore, we would expect to see emission
lines in thermal gas, and a Raymond-Smith model provides a very poor
fit.
The power law spectrum is steep with an energy spectral index 
\ax~=3.0 (see Table \ref{rass_fits}).
We also tried a
two power law model and a broken power law, but the same X-ray
spectral slopes were obtained in the soft and hard components. 
A single blackbody fit to the data does not give a reasonable fit, but using a
blackbody for the soft and a steep power law for the hard photons gives a good
fit (see Table \ref{rass_fits}).
For the power law model, we found excess
absorption of cold matter above the Galactic value. This explains the relatively
'hard' hardness ratio of $-0.20$ considering the steepness of the
spectrum.  While the steep power law (with intrinsic $\rm N_H$
cannot be ruled out
statistically,  it is a less favored model because it appears to be
difficult to produce physically.   

There are two pointed observations (ROR 141820 and 141829), obtained
on January 13, 1992, in which RX J1624.9+7554 could have been
detected. However, the source was detected in neither. We verified the
attitude solution using the bright star $\eta$ UMi that is visible in
the field ROR 141829.  We measured an upper limit for the count rate
of RX J1624.9+7554 of 0.0023 $\rm cts~s^{-1}$ at the expected
location.

\begin{figure}[t]
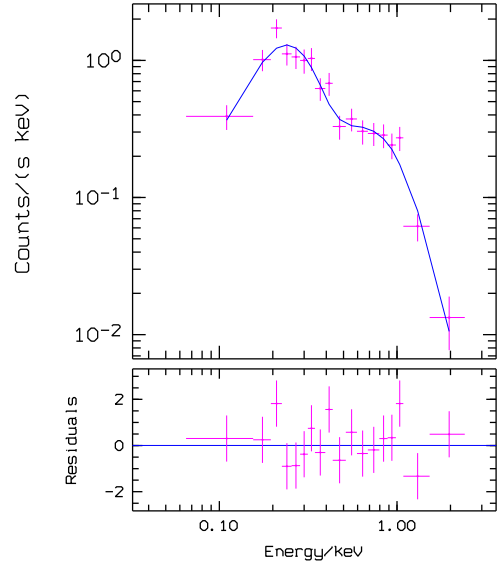

\clipfig{Bh041_F1}{67}{15}{90}{142}{182}
\clipfig{Bh041_F1}{67}{15}{15}{142}{65}
\caption[ ]{\label{rass-spec}
Power law fit to the RASS spectrum of RX J1624.9+7554. 
}
\end{figure}

\begin{table*}
\caption{\label{rass_fits} Spectral fits to the RASS spectrum of RX
J1624.9+7554. 
 ``$N_{\rm H}$'' is the column density given in 
units of $10^{21} \rm cm^{-2}$, ``Norm'' is the normalization 
at 1.0 keV (rest
frame) in $\rm 10^{-3}~Photons~cm^{-2}~s^{-1}~keV^{-1}$, 
``$\alpha$'' the  energy spectral slope,  ``A$_{\rm bb}$'' the black body
integral in $10^{-3}~ Photons~cm^{-2}~s^{-1}$, 
``A$_{rasm}$'' the normalization amplitude (in units if $10^{-3}$,
see EXSAS manual, Zimmermann et al. 1998),
and ``$T_{\rm rad}$'' and 
``$T{\rm plasma}$'' 
the radiation and plasma temperatures in eV. Models are power law (powl), 
blackbody (bbdy), thermal Bremsstrahlung (thbr), and Raymond-Smith thermal
plasma (rasm). All errors refer to the 1$\sigma$ level.
}
\begin{flushleft}
\begin{tabular}{lcccccccc}
\hline\noalign{\smallskip}
Model & $N_{\rm H}$ &
Norm & $\alpha_{\rm X}$ & $A_{\rm bb}$ & $A_{\rm rasm}$ & 
$T_{\rm rad}$ & $T_{\rm plasma}$ & $\chi^{2}/\nu$ \\
\noalign{\smallskip}\hline\noalign{\smallskip}
powl & 0.61\pl0.19  & 1.19\pl0.17 & 3.05\pl0.38 & --- & --- &
--- & --- & 15/15
\\
powl & 0.39 (fix)  & 1.15\pl0.14 & 2.29\pl0.12 & --- & --- & 
--- & --- & 22/16 \\
bbdy & 0.39 (fix)  & --- & ---& 35.4\pl3.8 & --- & 97\pl4 & --- & 50/16 \\
bbdy + powl & 0.39 (fix)  & 0.68\pl0.49 & 2.38\pl0.49 & 7.45\pl14.4 & ---   
& 107\pl58 & --- & 14/14 \\
thbr & 0.39 (fix)  & 1.14\pl0.17 & --- & --- & --- & --- & 283\pl21 & 17/16 \\
rasm & 0.39 (fix)  & --- & --- & --- & 4.37\pl0.35 & --- & 156\pl6 & 65/16 \\
\noalign{\smallskip}\hline\noalign{\smallskip}
\end{tabular}
\end{flushleft}
\end{table*}

\begin{figure}[h]
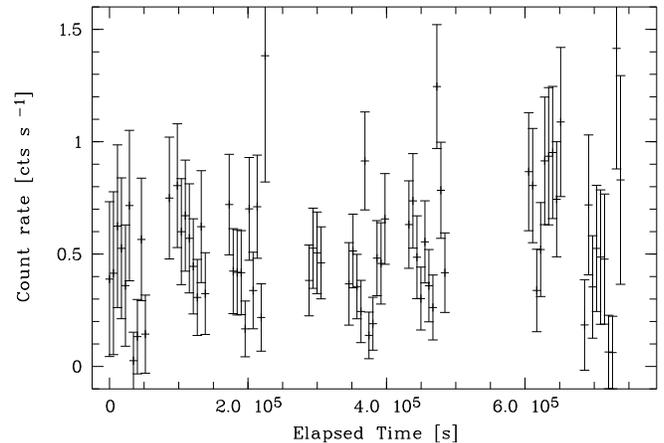

\clipfig{Bh041_F2}{87}{18}{21}{273}{195}
\caption[ ]{\label{lightcurve}
Light curve of RX J1624.9+7554 of the RASS data. The observation started October
07, 1990, 10:13. The errors refer to the 1$\sigma$ level.
}
\end{figure}

Fig. \ref{lightcurve} displays the RASS lightcurve of RX
J1624.9+7554 for all satellite orbits passing over it.
A variability test leads to a $\chi^2$ = 113 (71 dof) for a
constant hypothesis. The source is somewhat variable around a mean
count rate of 0.54  by a factor of about 2 on the  timescale of a day. 
The count rates of the sixth day are slightly higher than the rest.

\subsection{Optical}

\begin{figure}[t]
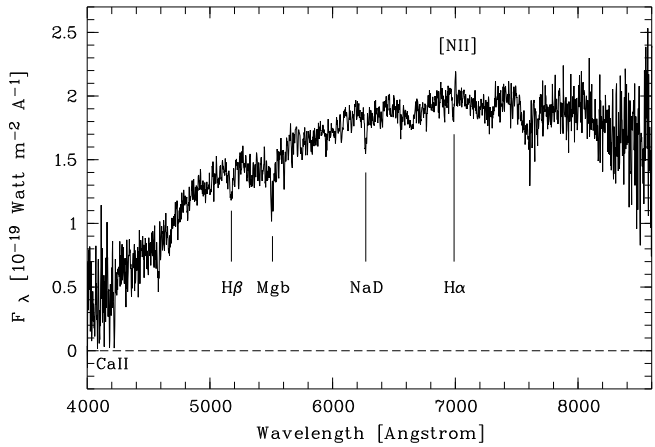

\clipfig{Bh041_F3}{87}{18}{10}{273}{195}
\caption[ ]{\label{mdmspec}
Combined
optical spectrum of RX J1624.9+7554 obtained with the 2.4m telescope at MDM
}
\end{figure}

The nuclear optical spectra from RX J1624.9+7554 taken in 1998 and 1999 are
typical of a non-active galaxy with some stellar absorption lines (see
Fig. \ref{mdmspec}). The only sign of activity is a weak
[NII]$\lambda$6584 emission line.  In the 2-D spectra extranuclear
H$\alpha$ and [NII] emission is seen, probably from HII regions.
The [NII] line in the nuclear spectrum may also be from an HII region.
A first look at the 1998 McD spectra
led to a probable mis-classification of the object as a BL Lac (Thomas et al.
1998).
From the $5.1^{''}$ slit 1998 McD spectrum
we measure magnitudes V = 16.2 and R=15.7. From the R-band image a magnitude of
15.6 was obtained.
The optical polarization measurements yield a degree of polarization of
0.35\pl0.31\%. Therefore, the source is essentially unpolarized.

\subsection{Spectral Energy Distribution}

\begin{figure}[t]
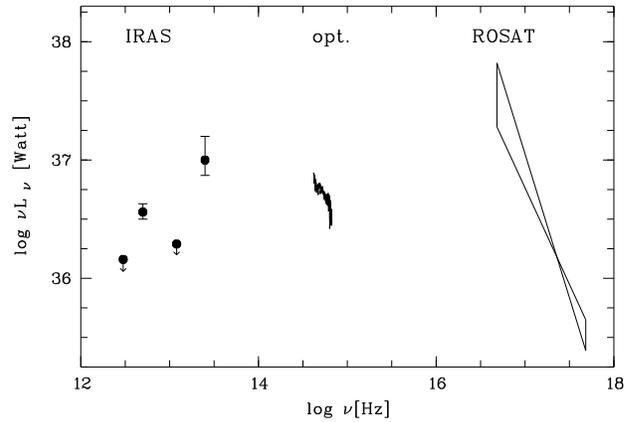

\clipfig{Bh041_F4}{87}{10}{10}{280}{195}
\caption[ ]{\label{sed_plot}
Spectral Energy Distribution of RX J1624.9+7554. The optical data are the 
broad slit McD spectrum.
The infrared data are IRAS scans and the X-ray data are represented by a error
bow-tie from the RASS observation and a power law fit. 
All data shown are in the observer's frame. 
}
\end{figure}

Fig. \ref{sed_plot} shows the Spectral Energy Distribution (SED)
of RX J1624.9+7554. The optical data are the 1998 wide-slit McD spectrum.
The X-ray spectrum is  represented by a power law 
model with free absorption parameter $N_H$
from the RASS observation with the limits given in Table \ref{rass_fits}.
Infrared data yield detections of 50 and 90 mJy at 12 and 60 $\mu$m,
respectively.

We looked for radio data and found that the nearest source in the 
NRAO VLA Sky Survey (NVSS) catalog is 6.73 arcminutes away from our
source.  This catalog contains objects with fluxes stronger than 
 $S~\approx~2.5$ mJy (Condon
et al. 1998).  

\section{\label{discussion} Discussion}
\subsection{The nature of RX J1624.9+7554} 
RX J1624.9+7554 is one of a few X-ray transient sources for which the X-ray
emission  vanished  between the RASS and  pointed observations.
We found an intrinsic 0.2-2.0 keV luminosity in the rest frame
of log $L_{X}$ [W] $=37.2\pm0.2$ using the parameters from the 
best-fitting absorbed power law model.
This high luminosity excludes the possibility of emission from a
highly luminous X-ray binary in the host galaxy.  It also excludes an
extremely bright supernova. First, in the last 100
years no supernova has been reported around the position of RX J1624.9+7554 and
second, the brightest supernovae have luminosities in the order of log L [W] 
= 34. Gamma Ray Burst (GRB) 
afterglows may be excluded because of the long relatively
constant X-ray light curve of RX J1624.9+7554 over a period of more than a week
during the RASS. 
There are several arguments against RX J1624.9+7554 
being a BL Lac. First, we do not find radio emission for this source. Second,
it is essentially unpolarized, third it is a spiral galaxy and so far BL Lacs
have almost
always been reported to be hosted by elliptical galaxies 
(e.g. Urry et al 1999), and fourth 
the X-ray spectrum would be at the very steep end 
if RX J1624.9+7554 is a
BL Lac (see e.g. Greiner et al. 1996, Perlman et al. 1996, Urry et al. 1996,
Lamer et al. 1996 for comparison).
Each argument for itself does not exclude the possibility that it is a BL Lac.
However, considered all together it becomes very unlikely that RX J1624.9+7554 
is a BL Lac.
Another, rather simple, explanation for the vanishing of
the X-ray source RX J1624.9+7554 could be a big cloud of cold
absorbing gas in the line of sight. However, in this case we would expect signs
of activity in the galaxy, which we do not see

Why do we consider this source to be an AGN even though it does not show
any signs of activity in its optical spectrum? The source has shown a
dramatic turn-off in X-rays on a timescale of less than two years.
Variability on such a short timescale would be impossible in an
ordinary galaxy simply because of the large extent.
The X-ray event must have happened in a very small region
that can produce both a high X-ray luminosity. 
The only machine that fulfills those constraints would be an
AGN engine. On the first view, the vanishing of 
RX J1624.9+7554 in X-rays appears similar to the case of 
WPVS007 (Grupe et al. 1995b); we interpreted that to be due to a shift
of the soft X-ray spectrum out of
the ROSAT PSPC energy window. However, WPVS007 is an active
Narrow-Line Seyfert 1 galaxy and RX
J1624.9+7554 does not show any signs of nuclear 
activity, at least not in its optical spectra
from 1998 and 1999. A possible explanation of the dramatic X-ray event can be
the tidal disruption of a star by the central black hole.

\subsection{Tidal disruption of a star}
The X-ray results of RX J1624.9+7554 can be caused by
an X-ray outburst similar to that seen in IC 3599 (Brandt et al. 1995, Grupe et al.
1995a) or NGC 5905 (Komossa \& Bade 1999). In both cases, a tidal disruption of
a star by the central black hole is considered to be a likely cause of
the outburst (see
Komossa \& Bade and references therein). Similar to RX J1624.9+7554,
the optical spectrum from NGC 5905 does not show nuclear activity. 
A tidal disruption of a star can occur if a star orbiting  a
supermassive 
 black hole is disrupted by the gravitational field of the black hole.
Part of the debris will orbit and part will fall into in black hole. This
will produce an X-ray outburst such as seen for example in IC 3599.
Rees (1990) 
estimated that statistically every 10000 years such a tidal disruption
event can happen around
a massive black hole in a galaxy. The estimated duration for a tidal disruption
of a star is on the order of one year for a star 'eaten' by a $\rm 10^{6}~
M_{\odot}$ black hole. 
However, in this case we would expect signs of activity in the
galaxy, which we do not see. 
Tidal disruption of a star also explains why an X-ray
outburst can be seen in a non-active galaxy.   
Outbursts can also
potentially come from instabilities in an accretion disk; however,
again in this case we would expect to see signs of activity in the
optical spectrum. Meanwhile, another X-ray outburst has been discovered in 
a non-active galaxy, RX J1242.6--1119  (Komossa \& Greiner 1999).

Our study of RX J1624.9+7554 is lacking one aspect: we do not have
simultaneous optical and X-ray data.  Therefore, unfortunately we do
not know what the optical spectrum looked like during the X-ray
outburst. In the case of IC 3599, we were lucky that optical
observations were made about half a year after the RASS (Brandt et al.
1995). It is important for future missions to perform 
repeated surveys,
such as it was planned for ABRIXAS. In this way we would be
able to detect activity and react much faster than we were able to in the
case of RX J1624.9+7554.

\acknowledgements{ 
We would like 
to thank I. Yadigaroglu for taking the R-image,
Dr. Derek Wills for measuring the polarization of RX J1624.9+7554, and   
Drs. Stefanie Komossa, Sally Laurent-Muehleisen, Jules Halpern, Thomas Boller,
and Stefan Immler for useful and intensive discussions. We would also 
like to thank our referee Dr. Eric S. Perlman
for his helpful comments and suggestions on this paper.
This research has made use of the NASA/IPAC Extragalactic
Database (NED).
We also
used the IRAS data request of the Infrared Processing and Analysis Center 
(IPAC), Caltech. 
KML gratefully acknowledges support through NAG5-7971 (NASA LTSA).
The ROSAT project is supported by the Bundesministerium f\"ur Bildung
und  Forschung (BMBF/DLR) and the Max-Planck-Society.

This paper can be retrieved via WWW: \\
http://www.xray.mpe.mpg.de/$\sim$dgrupe/research/refereed.html}

   \end{document}